# DESIGNING HIGH-SPEED, LOW-POWER FULL ADDER CELLS BASED ON CARBON NANOTUBE TECHNOLOGY

Mehdi Masoudi[1], Milad Mazaheri[1], Aliakbar Rezaei[1] and Keivan Navi[4]

[1]Department of Computer Engineering,
Science and Research Branch of IAU, Tehran, Iran
[4]Department of Electrical and Computer Engineering,
Shahid Beheshti University, Tehran, Iran

## ABSTRACT

*This article presents novel high speed and low power full adder cells based on carbon nanotube field effect transistor (CNFET). Four full adder cells are proposed in this article. First one (named CN9P4G) and second one (CN9P8GBUFF) utilizes 13 and 17 CNFETs respectively. Third design that we named CN10PFS uses only 10 transistors and is full swing. Finally, CN8P10G uses 18 transistors and divided into two modules, causing Sum and Cout signals are produced in a parallel manner. All inputs have been used straight, without inverting. These designs also used the special feature of CNFET that is controlling the threshold voltage by adjusting the diameters of CNFETs to achieve the best performance and right voltage levels. All simulation performed using Synopsys HSPICE software and the proposed designs are compared to other classical and modern CMOS and CNFET-based full adder cells in terms of delay, power consumption and power delay product.*

## KEYWORDS

*Full Adder, CNT, Carbon Nanotube Field Effect Transistor, High Speed, Low Power, High Performance & Power Delay Product*

## 1. INTRODUCTION

In digital electronic world, delay and power consumption improvement are the most important performance parameters of a circuit. To reach this goal, we can reduce scaling of the feature size. In complementary metal oxide semiconductor (CMOS) technology, reducing the length of channel to below about 65nm leads to critical problems and challenges such as decreasing gate control, short channel effect, high power density, high sensitivity to process variation and exponential leakage current increment [1]. For this reasons reducing the transistors size finally will stop at a point, leading to taking advantage of new technologies that do not have above problems may be felt. Therefore, new technologies such as benzene rings, quantum dot cellular automata (QCA), single electron transistor (SET), carbon nanotube field effect transistor (CNFET) and others have risen up [2-7].

Special properties of the carbon nanotubes (CNTs) cause to be utilized in various industries such as nanoelectronic. Some of these features are high thermal conductivity, high tensile strength, super conductivity, extreme rigidity and be conductor or semiconductor basis on structure.

Carbon nanotube field effect transistor (CNFET) is an appropriate alternative for CMOS. Owing to similarity between CNFET and CMOS in case of operation principle and the device structure, we can perform the established CMOS design infrastructure and CMOS manufacturing process in the CNFET technology [2]. CNFET is one of the molecular devices that avoid most fundamental silicon transistor restrictions and have ballistic or near ballistic transport in its channel and have high current carry ability [6-8]. Generally CNFET is faster and use lesser power consumption compared to silicon-based MOSFET and for this reason it is very suitable for high frequency and low voltage applications. Another useful trait of CNFET is that P-CNFET and N-CNFET have the same mobility and the same current drive capability, which is very important for transistor sizing in the complex circuits [9].

Recently some designs have been performed by CNFETs such as Galois field circuits [10], multiple valued logic circuits [1, 10, 11], interconnection networks [12-14] and CNFET full adders [13, 15]. Full adder cell is a basic element for complex circuits also most of arithmetic operations can be implemented by full adder cells. So performance increment of full adder causes to improve system performance [16]. Many efforts have been done to increase the efficiency of this element. Full adder performance improvement can be performed in various stages such as algorithm, circuit and technology. In this paper we present design details of high speed and low power full adder cells based on nanotechnology.

The rest of this paper is organized as follows. In section 2 a brief description about CNFET is provided. Previous works are presented in section 3. The proposed full adder cells are presented in section 4. Experimental results, analyses and comparisons are presented in section 5 and finally section 6 concludes the paper.

## 2. CARBON NANOTUBE FIELD EFFECT TRANSISTORS (CNFETS)

Carbon nanotube in theoretical definition is a sheet of graphite which is rolled up along a wrapping vector [17]. CNT can be single-wall (SWCNT) or multi-wall (MWCNT) [18]. Single-wall CNT is made by one layer of graphite and multi-wall CNT is made by more than one layer of graphite, then it is rolled up and all cylinders center is same. Carbon nanotubes are represented by a vector is called chirality vector. According to Figure 1, chirality vector is shown by $\vec{C}_n$ and is obtained by: $\vec{C}_n = n_1 \cdot \vec{a_1} + n_2 \cdot \vec{a_2}$, where $\vec{a_1}, \vec{a_2}$ are lattice unit vectors and $n_1, n_2$ are positive integers which specify the tube's structure [19]. Depending on $n_1$ and $n_2$, we have three different kinds of nanotube: zigzag, armchair and chiral (Figure 1(a)) and the SWCNT has different manners such that if $n_1 - n_2 = 3k \, (k \in \mathbb{Z})$ then SWCNT will be conductor otherwise SWCNT will be semiconductor [6, 20]. Conductive CNT is used as nanowires and semiconductive CNT is applied as transistor channel [6]. In CNFETs, one or more semiconductive SWCNTs can be used and a typical CNFET is illustrated in Figure 1(b).

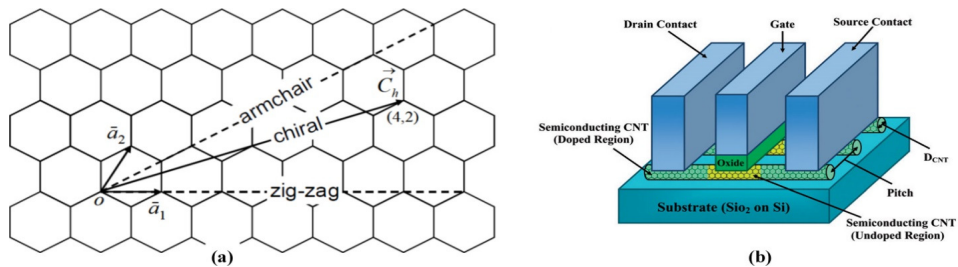

Figure 1. (a) Representation of a SWNT by a chiral vector [8], (b) schematic of CNFET [21]

$W_{gate}$ of CNFET is calculated approximately based on following equation: $W_{gate} \approx \text{Min}(W_{min}, N \times \text{pitch})$ [22]. where pitch is the distance between the centers of two neighbor SWCNTs under the same gate, $W_{min}$ is the minimum width of the gate and N is the number of nanotubes.

The diameter of nanotube $(D_{CNT})$ is given by Equation (1) [22].

$$D_{CNT} = \frac{0.249\sqrt{n_1^2 + n_2^2 + n_1 n_2}}{\pi} \qquad (1)$$

The threshold voltage of CNFET is an inverse function of the diameter of CNT. We use this ability to control $V_{th}$ by changing the $D_{CNT}$. The CNFET threshold voltage ($V_{th}$) is calculated based on Equation (2) [22].

$$V_{th} \approx \frac{E_g}{2.e} \quad , \quad E_g = \frac{2\sqrt{3}}{3}\frac{a.V_\pi}{D_{cnt}}$$
$$\Rightarrow V_{th} \approx \frac{\sqrt{3}}{3}\frac{a.V_\pi}{eD_{cnt}} \approx \frac{0.43}{D_{cnt}(nm)} \qquad (2)$$

Where parameter $E_g$ is the band gap energy, parameter a ($\approx 0.249$ nm) is the carbon-to-carbon atom distance, $V_\pi$ is ($\approx 3.033$ eV) the carbon $\pi$-$\pi$ bond energy, e is the unit electron charge and $D_{CNT}$ is the diameter of CNT.

Depending on the type of connections between source and drain with CNT channel and type of source, drain and gate, there are three main CNFETs. The first type is schottky barrier CNFET (SB-CNFET), which the CNT directly contacts to metal source. Schottky barrier junction limits the transconductance in the ON state, thus ION/IOFF ratio becomes rather low. SB-CNFET is appropriate for medium to high-performance applications. The second type of CNFET is the band-to-band tunneling CNFET (T-CNFET). T-CNFET has super cut-off characteristics and low ON current. These specifications make it suitable for low power and subthreshold application but it is not appropriate for very high speed applications. The third type of CNFET is MOSFET-like CNFETs. Source and drain are doped with positive impurities, so a semiconductor-semiconductor junction between source-drain and channel is made and source-channel junction is not schottky barrier. As a result MOSFET-like CNFETs have high ON current, high ION/IOFF ratio and scalability that make it suitable for high performance applications [6, 23, 24]. In this paper we employ MOSFET-like CNFETs for all proposed designs.

## 3. PREVIOUS WORKS

As mentioned before, CNFET is a suitable alternative for silicon-based technology that does not have MOSFET problems. Many full adder designs by MOSFET and CNFET are proposed so far and each of them has their advantages and disadvantages. In this paper we select 5 popular full adder cells that three of them are designed with 32nm MOSFET technology and two of them designed with 32nm CNFET technology. In this section a brief description of some prior works is presented.

First one is conventional CMOS full adder (CCMOS). It has 28 transistors and consumes high power and area[25, 26]. This design is based on standard CMOS topology and has full swing output that increases noise margin and reliability. Because of using high number of PMOS in pull up network, this design has high input capacitance, leading to high delay and dynamic power consumption. Although, using inverters on the output nodes decreases the rise-time and fall-time delay and increases the driving ability. Next sample full adder cell is TG-CMOS which has 20 transistors [27]. This full adder uses conventional transmission gates on its structure. In TG-CMOS full adder all outputs are obtained based on XOR/XNOR gates and afterwards uses MUX structure. This design has low power dissipation because of using transmission gate but utilizes high number of transistors.

Next design, CMOS-Bridge uses 24 transistors and takes advantage of the high-performance bridge structure [28]. In this design Cout signal is produced based on a CMOS style and Sum signal is generated from Cout by means of a bridge circuit. In addition, to generate Cout and Sum from $\overline{Cout}$ and $\overline{Sum}$ and for enhancing the driving capability, two inverters are utilized at the output nodes.

Next full adder is proposed in [29] and has been designed by CNFETs (Figure 2(a)) (we named it CNT-FA1 in this paper). It uses 14 transistors plus three capacitors. This design is based on majority function and $V_{th}$ is adjusted by changing the diameters of CNFETs. CNT-FA1 provides good delay and power properties.

Full adder cell that proposed in [30] (we named it CNT-FA2 in this paper) uses majority function plus inverter to generate $\overline{cout}$ and also by meaning to adjusting the $D_{CNT}$ and as result, adjusting the Vth in PCNFET and NCNFET, NAND and NOR has been created by inverter structure (Figure 2(b)). This design uses two pass transistors after NAND and NOR to generate $\overline{SUM}$ output and using pass transistor causes to the output does not be full swing but the final inverter fixes it and makes full swing output. This design is symmetric and delay, power consumption and PDP parameters are suitable.

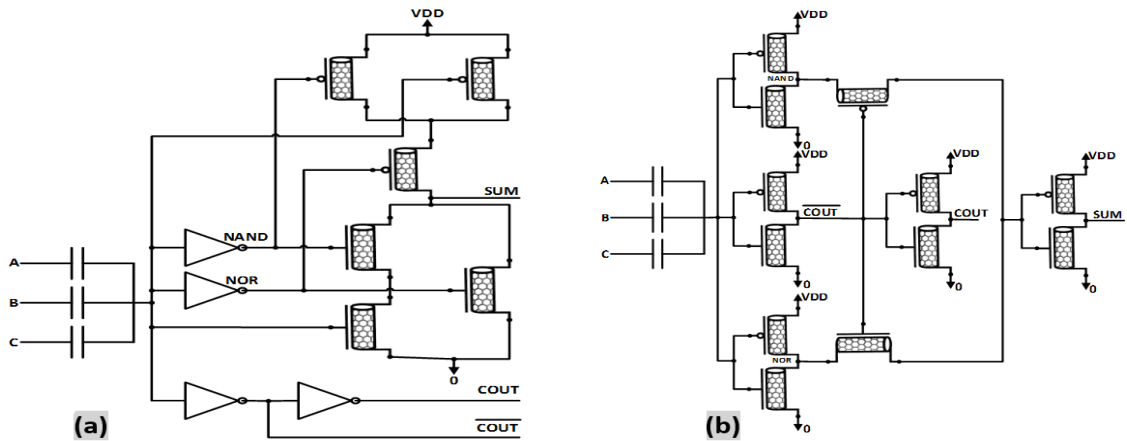

Figure 2. (a) CNT-FA1 full adder, (b) CNT-FA2 full adder

## 4. PROPOSED DESIGNS

In this section we propose four new full adder cells. The logic formula for a one-bit full adder is shown in Equation (3). The inputs are A, B, C (C is carry input) and outputs are Sum and Cout. XOR is shown by $\oplus$ symbol.

$$SUM = XOR(A,B,C) = A \oplus B \oplus C \quad (3)$$
$$= (A \oplus B) \oplus C \quad , \quad A \oplus B = A\bar{B} + \bar{A}B$$
$$COUT = A.B.C + A.B.\bar{C} + A.\bar{B}.C + \bar{A}.B.C$$
$$= A.B + C(A.\bar{B} + \bar{A}.B) \quad (4)$$
$$= A.B + C(A \oplus B)$$

According to Equation (3), we can generate Sum by using XOR twice. First, A and B become XOR and then the result of previous stage become XOR with C. A XOR module is illustrated in Figure 3. This module uses two pass transistors and two pull down transistors. Pass transistors can cover the 3 states of inputs that are 00, 01, 10 and one more remained state (11) is handled by pull down network. By using twice of this module, $(A \oplus B) \oplus C$ is obtained. All proposed designs use this module to generate Sum signal.

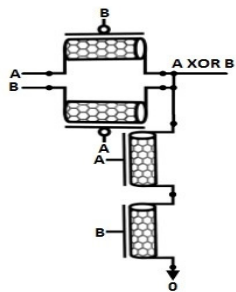

Figure 3. XOR module

According to Equation (4), we can use $A \oplus B$ to generate Cout. So in the first proposed design, we have used a transmission gate to lead C into Cout when $A \oplus B$ is high. In this status two states remain, when both A and B are high or both A and B are low. First one covered by a pair series of NCNFETs and second one covered by another pair series of PCNFETs. This design uses 13 transistors. We named it CN9P4G and shown in Figure 4.

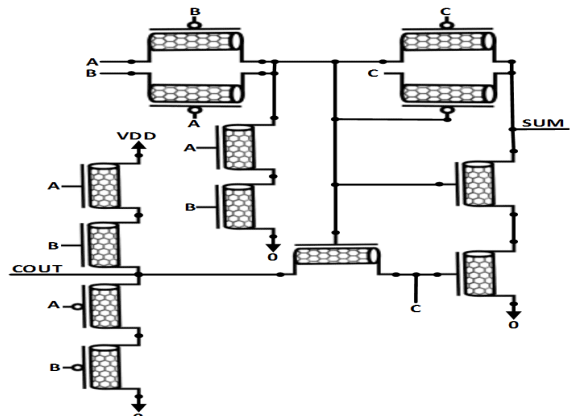

Figure 4. CN9P4G (Proposed1)

This design and others that will present later, depending on input values can have threshold losing in output voltages and are not full swing. For solving this problem we have some method such as using transmission gate instead of pass transistor or using output buffers, but these methods cause to increase transistor using and critical path and in result increasing power consumption and delay. CNFETs have a special property that can be used for this case. According to Equation (2), the threshold voltage is inversely proportional to the diameter of its CNT. So by increasing the diameter of CNT, $V_{th}$ can be reduced. Decreasing of $V_{th}$ leads to better driving capability and higher speed and the full swing problem can be solved. According to Equation (1), $D_{CNT}$ is obtained by chiral vector with $(n_1, n_2)$ pair.

So as a result we use $(n_1, n_2) = (55,0), D_{CNT} = 4.306nm$ for all PCNFETs and NCNFETs. By this amount of diameter, $V_{th}$ will be very low.

According to simulation results, voltage drop is very lower than the value of the threshold. This is due to the very high-speed operation of CNFETs with large diameters in subthreshold region. Larger CNT diameter leads to have smaller band gap, and smaller band gap leads to higher on-currents and shorter propagation delay. In addition, CNFETs with smaller band gap and as result smaller threshold voltage, are less sensitive to process variations and it leads to better manufacturability [31].

In first proposed design (CN9P4G) two states are exist that Cout is not full swing, when ABC=001 or ABC=110. When these input patterns occur, pass transistors will have current in subthreshold region and causes to destroyed output voltage. To fix this problem we use a buffer in Cout output in second proposed design, CN9P8GBUFF (Figure 5). Using buffer causes to increase critical path to 5 transistors and consequently more power consumption and delay rather than first design (CN9P4G). Transistors using is also increased to 17, but outputs become full swing.

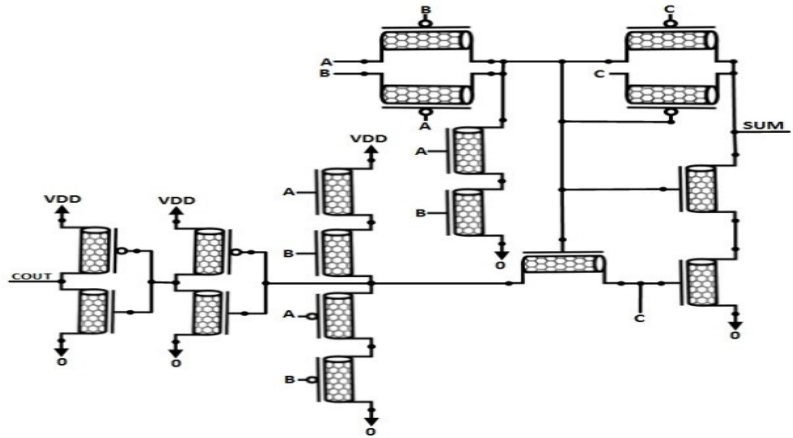

Figure 5. CN9P8GBUFF (Proposed2)

In proposed design 3, Cout is generated in another way and transistor using decrement is also significant (Figure 6). In this design we implement Equation (5) by a NCNFET and a PCNFET pass transistor.

$$\begin{aligned} COUT &= C(A \oplus B) + AB \\ &= C(A \oplus B) + A\bar{A}\bar{B} + AAB \\ &= C(A \oplus B) + A(\bar{A}\bar{B} + AB) \\ &= C(A \oplus B) + A(\overline{A \oplus B}) \end{aligned} \quad (5)$$

Total used transistors are 10. This design due to using large diameter CNFETs $(D_{CNT} = 4.306nm)$ is full swing.

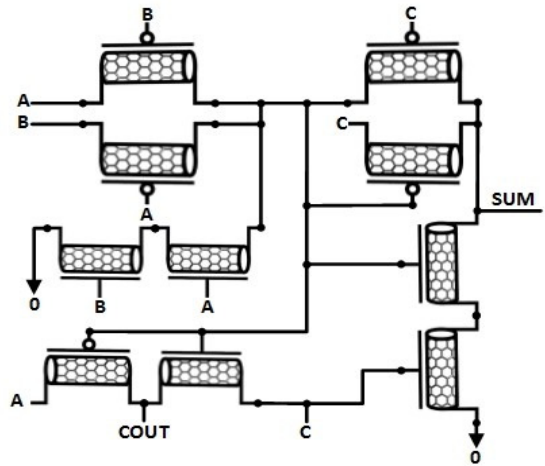

Figure 6. CN10PFS (Proposed3)

Next design (CN8P10G) includes two modules to generate Sum and Cout signals separately (Figure 7). CN8P10G has a parallel manner in producing Sum and Cout and uses 18 transistors.

Sum made by using XOR module (Figure 3) twice like other proposed designs. Cout signal has been generated by using Equation (6). Also all output voltages are full swing.

$$COUT = Majority(A, B, C)$$
$$= A.B + A.C + B.C \quad (6)$$

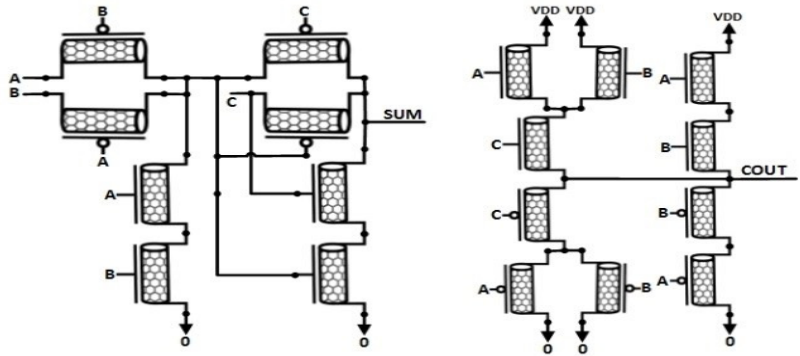

Figure 7. CN8P10G (Proposed4)

## 5. SIMULATION RESULTS ANALYSIS AND COMPARISON

All proposed designs and previous designs were described in section 3 and 4, have been simulated by Synopsys HSPICE 2010. CMOS-based circuits are simulated using 32nm CMOS technology and CNFET-based circuits are simulated using compact SPICE model for 32nm [8, 32-34]. Compact SPICE model has been designed for unipolar MOSFET-like CNFET devices and each transistor can have one or more CNTs. This model considered schottky barrier effects, parasitics, CNT charge screening effects, CNT Source/Drain and gate resistances and capacitances. The parameters of this CNFET model, corresponding values and a brief description are presented in Table 1.

Table 1. CNFET model parameters

| Parameter | Description | Value |
|---|---|---|
| Lch | Physical channel length | 32nm |
| Lgeff | The mean free path in the intrinsic CNT channel | 100nm |
| Lss | The length of doped CNT source-side extension region | 32nm |
| Ldd | The length of doped CNT drain-side extension region | 32nm |
| Kgate | The dielectric constant of high-k top gate dielectric material | 16 |
| Tox | The thickness of high-k top gate dielectric material | 4nm |
| Csub | The coupling capacitance between the channel region and the substrate | 40 pF/m |
| Efi | The fermi level of the doped S/D tube | 6 eV |

All four proposed designs output waves are shown in Figure 8. As can be seen CN9P4G is not full swing for 001 and 110 input patterns and all other designs are full swing.

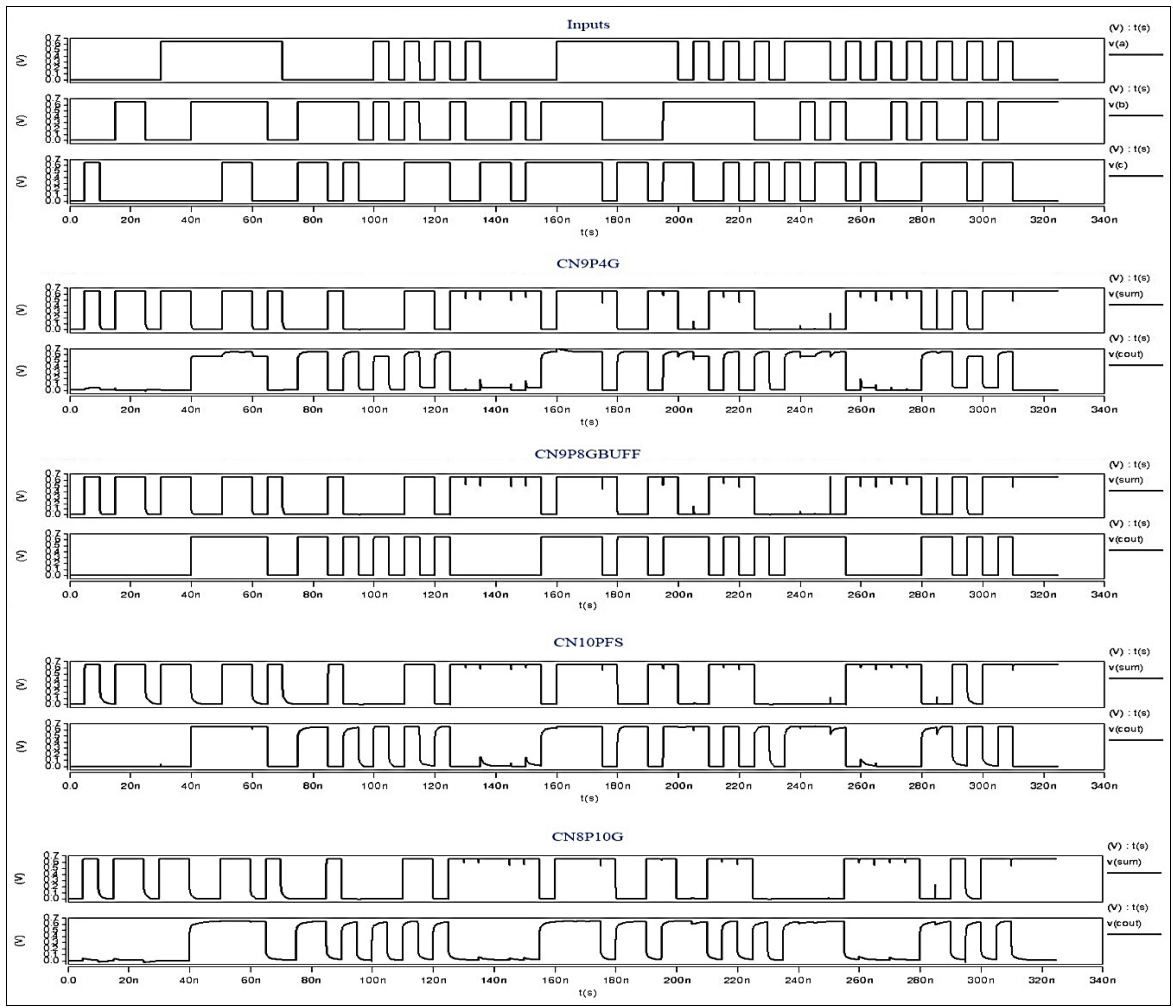

Figure 8. The input and output signals of 4 proposed full adder

Main comparison factor in this paper is PDP (Power Delay Product), so in variation of load capacitor, frequency, power voltage and temperature, all PDPs have been calculated.

Table 2 shows simulation results for all designs at 250 MHz frequency, 2.1 *femto Farad* load capacitor and 25° temperature in variation of supply voltage from 0.5v to 0.8v. Delay, power consumption and PDP of our designs have been shown in this Table. Our proposed full adders have lower delay, power and PDP rather than other mentioned previous circuits. For example, according to Table 2, PDP of CN8P10G at 2.1 *femto Farad* capacitor and power voltage 0.65 has 86.1% improvement rather than CMOS-bridge, 80.9% improvement rather than C-CMOS, 76.9% improvement rather than TG-CMOS, 79.3% improvement rather than CNT-FA1 and 66.7% improvement rather than CNT-FA2.

Table 2. Simulation results at frequency=250MHz and load capacitor=2.1 *femto Farad*

|  |  | Vdd=0.5 v | Vdd=0.65 v | Vdd=0.8 v |
|---|---|---|---|---|
| Delay (E-10 s) | CMOS-Bridge | 4.9264 | 1.926 | 12.002 |
|  | CCMOS | 3.9300 | 1.444 | 9.4001 |
|  | TG-CMOS | 2.3753 | 0.88103 | 5.4938 |
|  | CNT-FA1 | 3.0340 | 0.8747 | 4.8074 |
|  | CNT-FA2 | 2.1070 | 0.79694 | 5.8841 |
|  | CN9P4G(P1) | 0.40743 | 0.29928 | 0.27675 |
|  | CN9P8GBUFF(P2) | 0.43620 | 0.32865 | **0.27409** |
|  | CN10PFS(P3) | 0.45567 | 0.34379 | 0.28079 |
|  | CN8P10G(p4) | **0.27607** | **0.27331** | 0.86408 |
| Power (E-7 w) | CMOS-Bridge | 1.6830 | 3.0314 | 5.2361 |
|  | CCMOS | 1.6369 | 2.926 | 5.0607 |
|  | TG-CMOS | 2.1678 | 3.9688 | 7.6154 |
|  | CNT-FA1 | 1.5523 | 4.4724 | 3.1228 |
|  | CNT-FA2 | 1.3761 | 3.0466 | 1.6917 |
|  | CN9P4G(P1) | 1.9720 | 3.0276 | 4.3138 |
|  | CN9P8GBUFF(P2) | 2.1721 | 3.3545 | 4.9056 |
|  | CN10PFS(P3) | **1.8339** | **2.8507** | **4.1671** |
|  | CN8P10G(p4) | 1.8620 | 2.9572 | 4.3012 |
| PDP (E-17 J) | CMOS-Bridge | 8.2742 | 5.8384 | 6.2844 |
|  | CCMOS | 6.4329 | 4.2253 | 4.7571 |
|  | TG-CMOS | 5.1492 | 3.4966 | 4.1837 |
|  | CNT-FA1 | 4.7097 | 3.912 | 15.013 |
|  | CNT-FA2 | 2.8995 | 2.4279 | 9.9541 |
|  | CN9P4G(P1) | 0.80345 | 0.9061 | 1.1938 |
|  | CN9P8GBUFF(P2) | 0.94746 | 1.1024 | 1.3446 |
|  | CN10PFS(P3) | 0.83566 | 0.98004 | **1.1701** |
|  | CN8P10G(p4) | **0.51404** | **0.80823** | 3.7166 |

In the next evaluation, load capacitor is increased from 1.4 to 4.9 *femto Farad* and other parameters are fixed at Vdd=0.65V, frequency=250MHz and temperature=25°. All delays, powers and PDPs are calculated and are shown in Figure 9, Figure 10 and Figure 11 respectively.

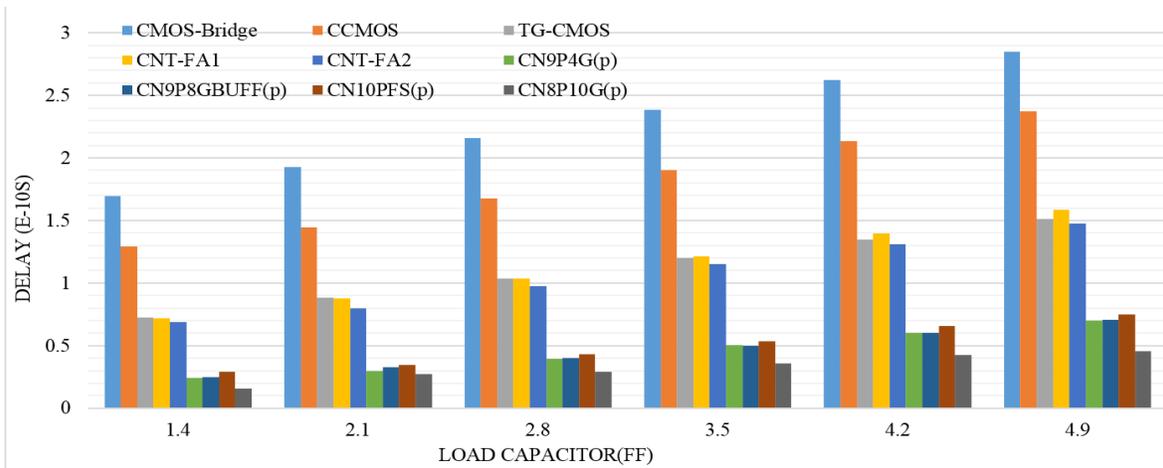

Figure 9. Delay of the circuits versus load capacitor variations

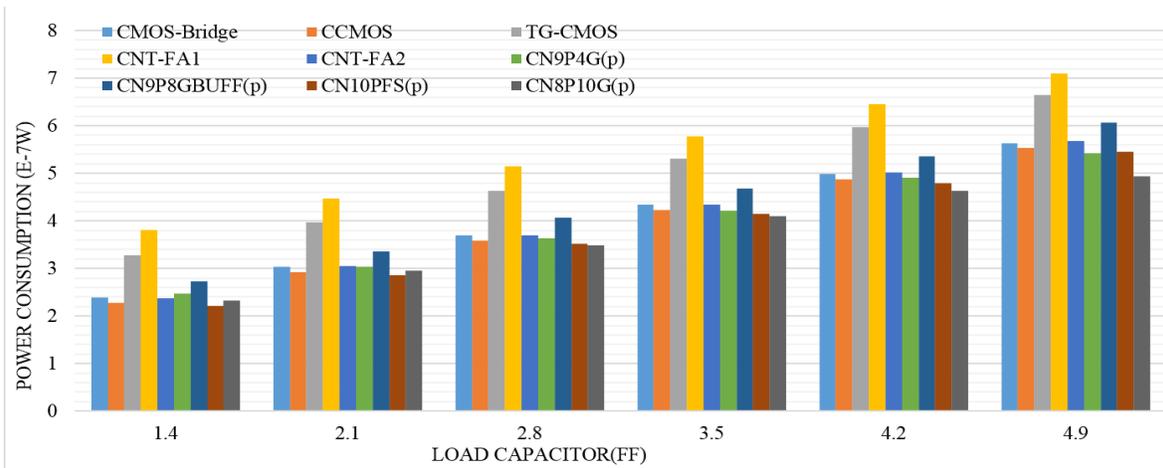

Figure 10. Power consumption of the circuits versus load capacitor variations

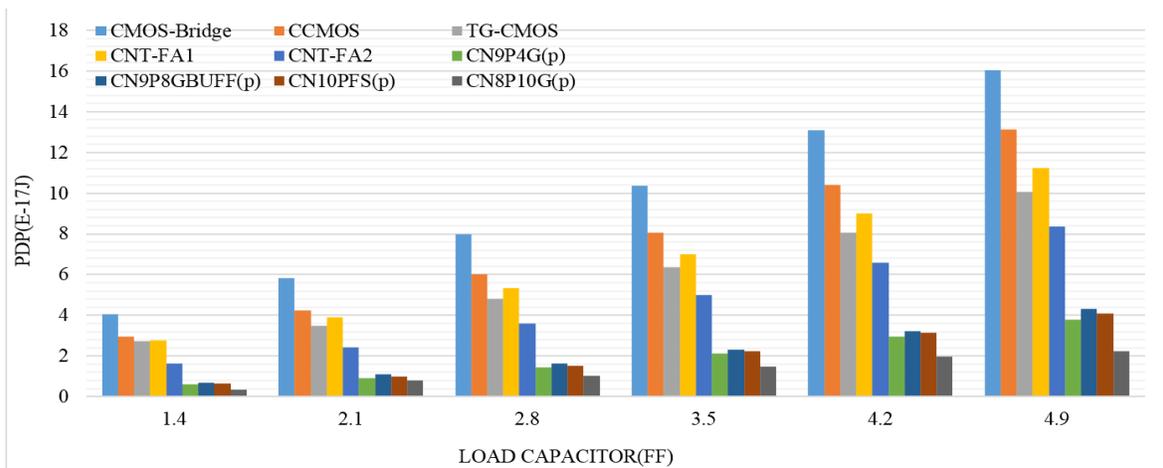

Figure 11. PDP of the circuits versus load capacitor variations

International Journal of VLSI design & Communication Systems (VLSICS) Vol.5, No.5, October 2014

Another changed parameter is frequency (Figure 12). This simulation has been performed in Vdd=0.65V, Cload=2.1 *femto Farad* and temperature=25°. CN8P10G has lower PDP in selected frequencies and also has lower increment ratio.

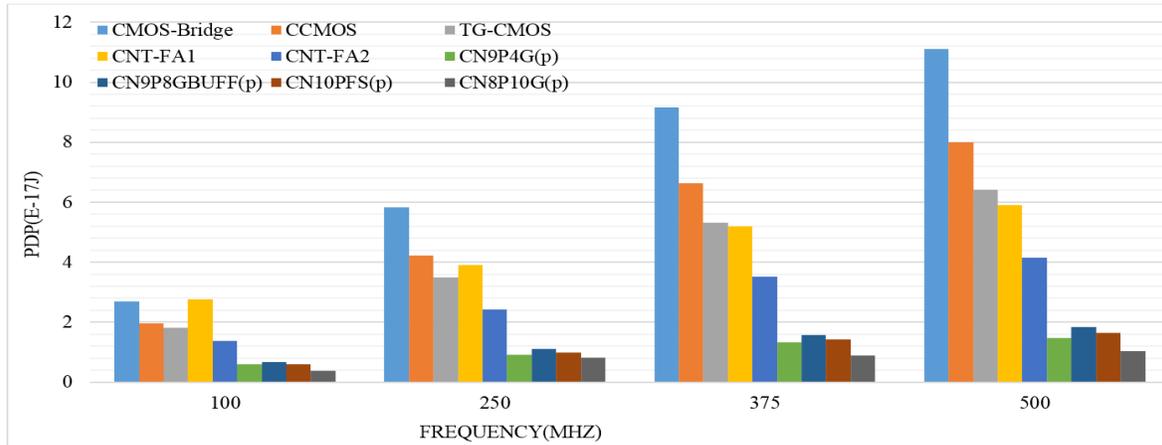

Figure 12. PDP of the circuits versus frequency variations

Figure 13 shows results with temperature variations from 0° to 70°, frequency=250MHz, load capacitor=2.1 *femto Farad* and power voltage=0.65V. PDP for all designs has been calculated. In this analysis, presented designs have lower PDP in all temperature.

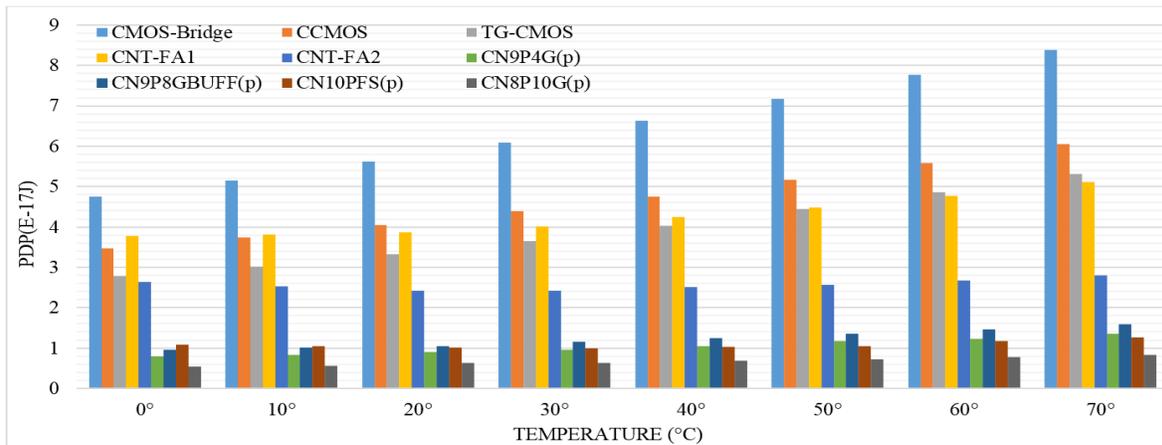

Figure 13. PDP of the circuits versus temperature variations

## 6. CONCLUSIONS

In this article some new full adder designs have been presented which are high speed, low power and high performance. By using exclusive properties of CNTFET, the efficiency of these designs was improved. Circuits were simulated in various conditions and simulation results approved the efficiency of proposed designs rather than other CMOS and CNFET-based designs that investigated before.

## AUTHORS

**Mehdi Masoudi** received his B.Sc. degree in hardware engineering from Shahid Beheshti University, Tehran, Iran, in 2007 and M.Sc. degree in computer architecture from IAU Science and Research branch, Tehran, Iran, in 2013. He is a research member at Nanotechnology and Quantum Computing Laboratory of Shahid Beheshti University. His research interests include High Performance VLSI Designs, Nanoelectronics and new emerging technologies specially CNFET.

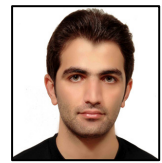

**Milad Mazaheri** received his M.Sc. degree in computer architecture from Science and Research Branch, IAU, Tehran, Iran in 2013. He is currently pursuing his Ph.D. degree in computer architecture from Science and Research Branch of IAU, Tehran, Iran. His research interests include Photonic Networks-on-chip, VLSI Systems Design, Embedded Systems, Multiprocessor Systems, Fault Tolerant Design and Computer Architecture.

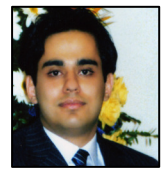

**Aliakbar Reazei** received his B.Sc. degree in hardware engineering from Shahid Beheshti University, Tehran, Iran, in 2007 and M.Sc. degree in computer architecture from IAU Science and Research branch, Tehran, Iran, in 2013. He is a research member at Nanotechnology and Quantum Computing Laboratory of Shahid Beheshti University. His research interests include High Performance VLSI Designs, Sensor Networks and Computer Architecture.

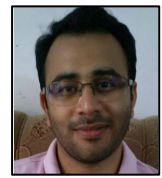

**Keivan Navi** received his M.Sc. degree in electronics engineering from Sharif University of Technology, Tehran, Iran in 1990. He also received his Ph.D. degree in computer architecture from Paris XI University, Paris, France in 1995. He is currently Professor in Faculty of Electrical and Computer Engineering of Shahid Beheshti University. His research interests include Nanoelectronics with emphasis on CNFET, QCA and SET, Interconnection Network Design, Quantum Computing and Cryptography. He has been a visiting professor at the University of California, Irvine.

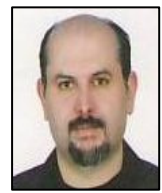